\documentclass[twocolumn,a4paper,10pt]{article}

\usepackage[british]{babel}        
\usepackage{graphicx}              %
\usepackage{hyperref}              
\usepackage{xspace}
\usepackage{tabularx}
\usepackage{subfigure}

\newcommand{\peershare}{\texttt{PeerShare}\xspace}

\hypersetup{
  pdfauthor={Marcin Nagy, N. Asokan, and Joerg Ott},
  pdftitle={\peershare: A System Secure Distribution of Sensitive Data Among Social Contacts}
  pdfsubject={Secure Data Distribution},
  pdfkeywords={Data distribution} {social networks} {access control}
}

\begin{document}
\title{\peershare: A System Secure Distribution of Sensitive Data Among Social Contacts}
\author{Marcin Nagy \\Aalto University     \\marcin.nagy@aalto.fi
  \and N. Asokan \\University of Helsinki \\ asokan@acm.org
  \and J\"org Ott   \\Aalto University     \\jorg.ott@aalto.fi
}
\date{15 July 2013}
\maketitle
\begin{abstract}
We present the design and implementation of the \peershare, a system that can be used by applications to securely distribute sensitive data to social contacts of a user.  \peershare incorporates a generic framework that allows different applications to distribute data with different security requirements.  By using interfaces available from existing popular social networks, \peershare is designed to be easy to use for both end users as well as developers of applications. 
\peershare can be used to distribute shared keys, public keys and any other data that need to be distributed with authenticity and confidentiality guarantees to an authorized set of recipients, specified in terms of social relationships.
We have used \peershare already in three different applications and plan to make it available for developers.
\end{abstract}

\section{Motivation}
\label{:motivation}
Key management has been one of the challenging problems in guaranteeing  secure communication between parties on the Internet. Although \emph{public key technology} holds the promise of simplifying key management, multiple technical, economic, legal and social reasons prevented PKIs to be successfully deployed in the Internet, thus leaving the problem of secure key distribution still wide open \cite{DBLP:journals/computer/Gutmann02}.

Recent years have also brought a tremendous increase in the popularity of social networks. Social networks (like Facebook\cite{Facebook}, or Twitter\cite{Twitter}), by giving users opportunity to share data among their friends and by providing APIs for third party developers, open new possibilities for creation of applications using social graph data. The most important aspects of social networks in the context of data distribution are:
\begin{itemize}
\item the possibility of common user authentication by means of the \emph{Single Sign-on} service and the \emph{OAuth} protocol \cite{rfc6749},
\item the extensive scale of deployment of popular social networks, and
\item the ability for users to express social relationships in an intuitive manner (e.g., ``friends'', ``colleagues'', ``friends of friends'' etc.)
\end{itemize}

Our starting point is the observation that one can use social networks to facilitate the distribution of authentic public keys~\cite{socialkeys}.  One can generalize this to design a generic framework that allows distribution of arbitrary application-specific sensitive data with specific security requirements (like authenticity-only or authenticity and confidentiality) to a specific set of social contacts.  The result is a system we call \peershare, which we describe in this paper.

\textbf{Our goal and contribution.} In this paper we present \peershare: the design and implementation of a system that allows users to distribute data securely. \peershare distinguishes itself from other data distribution systems through:
\begin{itemize}
\item incorporating \textbf{a generic framework for data distribution} that can be used by different applications to distribute different types of data (e.g., shared secret keys, public keys, other sensitive data) to a specified set of social contacts with different security guarantees. 
\item improving \textbf{usability both for end-users and application developers} by taking advantage of existing and popular social network tools for the user authentication and distribution of data inside a specific social context.
\end{itemize}

In our implementation, we use on Facebook as the social network.  The social network server is used for user authentication and for users to define social groups either as pre-defined lists (like ``friends'' or ``friends-of-friends'') or custom lists. However, our system is generic and can use any social network that supports a single sign on (SSO) and authorization mechanism (like OAuth 2.0) and provides an interface for apps access  user's social graph information. Given the scale of social networks deployment, the SSO using the social network greatly increases the usability of user authentication. Social graph information is used only for obtaining user specific friend lists which can be used to specify access control for the data being distributed.

The social network server is not involved in actual data distribution; that is done through the \peershare server containing database with data to be distributed to specified users. \peershare client-side implementation, called the \peershare Service, is responsible for uploading new data to the server, deleting old data items, and periodically querying the server to check if there are any new data for them uploaded by other devices. \peershare Service exposes an API towards applications which allow them to make use of \peershare functionality.  The communication between the client and the server is via an authenticated secure channel. \peershare server is assumed to be a trusted entity. In Section~\ref{:validation}, we discuss ways of reducing this trust assumption.

\textbf{Outline.} We describe usage scenarios in Section \ref{:usecase} which motivate usage of \peershare. Section~\ref{:requirements} presents system requirements. Section~\ref{:design} includes detailed system design, while Section~\ref{:validation} describes security considerations of the system. Section~\ref{:related} presents the related work. Finally, section~\ref{:conclusion} concludes the paper.

\section{Usage scenarios}
\label{:usecase}
Currently \peershare is already used in three example applications, namely PeerSense~\cite{DBLP:conf/percom/GuptaMNAW12}, SCAMPI~\cite{DBLP:journals/ccr/PitkanenKOCPGPLTHMHS12} and CrowdShare~\cite{CrowdShare-techrep}\cite{acns2013}.


\emph{PeerSense} \cite{DBLP:conf/percom/GuptaMNAW12} is a service on a mobile device that senses the presence of nearby friends. Applications can query PeerSense for the set of nearby friends at any given moment(e.g. the camera application can query the set of nearby friends at the time when the shutter is pressed and attach this information as metadata to the resulting picture~\cite{Qin:2011:TSA:1999995.1999997}. PeerSense uses \peershare to distribute the binding between the Bluetooth Device Address (BDADDR) and the social network identifier of the device user to the set of users who the device user wants to be visible to. Although BDADDR itself is not secret, the binding is. Hence, the data share via \peershare needs secrecy and authenticity. At any given time a device is associated with at most one user, whereas a user may be associated with multiple devices. Thus, the data is device-specific.

The \emph{SCAMPI} platform~\cite{DBLP:journals/ccr/PitkanenKOCPGPLTHMHS12} allows mobile devices to communicate in the opportunistic network. Mobile devices discover themselves through multicasting their SCAMPI identifiers, which are hashes of their public keys. As such identifiers are not very meaningful to use, the SCAMPI platform uses \peershare for mapping the SCAMPI identifiers to social identifiers. Similarly to PeerSense, the exchanged data is a binding, thus it is private and also device-specific. 

In the \emph{CrowdShare} project~\cite{CrowdShare-techrep}\cite{acns2013}, devices in the network are able to share their resources with one another based on existing social relationships. CrowdShare presents privacy-preserving friend of friend finder service based on the private set intersection (PSI) algorithm \cite{DGT12}. The input to PSI consists of a set of ``bearer tokens''. A bearer token is generated by the device of a user and is distributed to all friends of that user. It serves as a capability for proving the friend relationship. The data shared using \peershare are the bearer tokens. They are user-specific and require both authenticity and integrity.

\begin{table*}[ht!]
\caption{Summary of existing \peershare use cases}
\label{usecase_summary}
\centering
\begin{tabular}{ | c | c | c | c | }
\hline
\textbf{Use case} & \textbf{Type of data} & \textbf{Security need} & \textbf{Specificity} \\ \hline
PeerSense & BDADDR:social-ID & private & device-specific \\ \hline
SCAMPI & SCAMPI-ID:social-ID & private & device-specific \\ \hline
FoF finder & bearer token & private & user-specific \\ \hline
Public key distribution & public key & public & user-specific \\ \hline
\end{tabular}
\end{table*}

Furthermore, CrowdShare can optionally make use of user-specific public keys. Distribution of public keys is done similarly to the SocialKeys project~\cite{socialkeys}, but using \peershare. The exchanged data would then be a binding between a public key and a social identifier. Thus, such binding is public and user-specific.

Table~\ref{usecase_summary} presents a short summary of existing \peershare use cases.

Finally, there are also possible situations in which, we are interested in making a binding between a data item and a specific user that does not use the \peershare system. To do this, we introduce also the notion of data binding type. If data is uploaded normally by the application, we call it a owner-asserted binding. However, if a user decides to add a binding for another user, such binding is called user-asserted, and is only visible to the user that has created it. An example of such a situation is present in the PeerSense application. A user can tag a device on the list of scanned devices and assign a name of his/her friend to it. Since there is no evidence for the correctness of such a user-asserted binding, it is not distributed using \peershare, but is still available via the \peershare API in the devices of the user who asserted the binding.

Given the common aspects of these different cases of secure data distribution, it is evident that designing a generic data distribution framework would improve ease of development, use and security.

\section{System requirements}
\label{:requirements}
\subsection{\peershare goals}
Our vision of a successful data distribution system sets three basic goals for it to fulfil:
\begin{enumerate}
\item \textbf{Data security} $\Rightarrow$ assurance of data security is the most critical requirement to convince users to the distribution system. 
\item \textbf{Usability} $\Rightarrow$ the system should be as easy to use as possible for users, thus necessary user interaction should be minimised.
\item \textbf{Deployability} $\Rightarrow$ the system should be scalable to allow various application developers to easily distribute their data though it.
\end{enumerate}

\subsection{Assumptions}
Our threat model assumes that each device has platform security that:
\begin{itemize}
\item isolates applications from one another (both during execution time and in terms of persistent storage),
\item allows a service on the device to learn a platform-specific identity of a calling application that wants to access the service.
\end{itemize}



\subsection{Threats}
We need to provide protection against the following threats: 
\begin{enumerate}
\item \textbf{Man-in-the-middle Attacks} $\Rightarrow$ any network devices that route messages between the mobile device and the server should not be able to act as man-in-the-middle that eavesdrops on or modifies messages. 
\item \textbf{Unauthorized Usage} $\Rightarrow$ only the person that has created the data item should be able to later modify or erase it. Furthermore, as data are created by applications that use the \peershare system, only the application that has created the particular data item should be able to access, modify or delete it. Finally, data should be distributed by the \peershare server only to users that are eligible to obtain them. 
\end{enumerate}

\begin{figure*}[!ht]
	\centering
	\includegraphics[scale=0.5, trim=2cm 3cm 0cm 3.0cm]{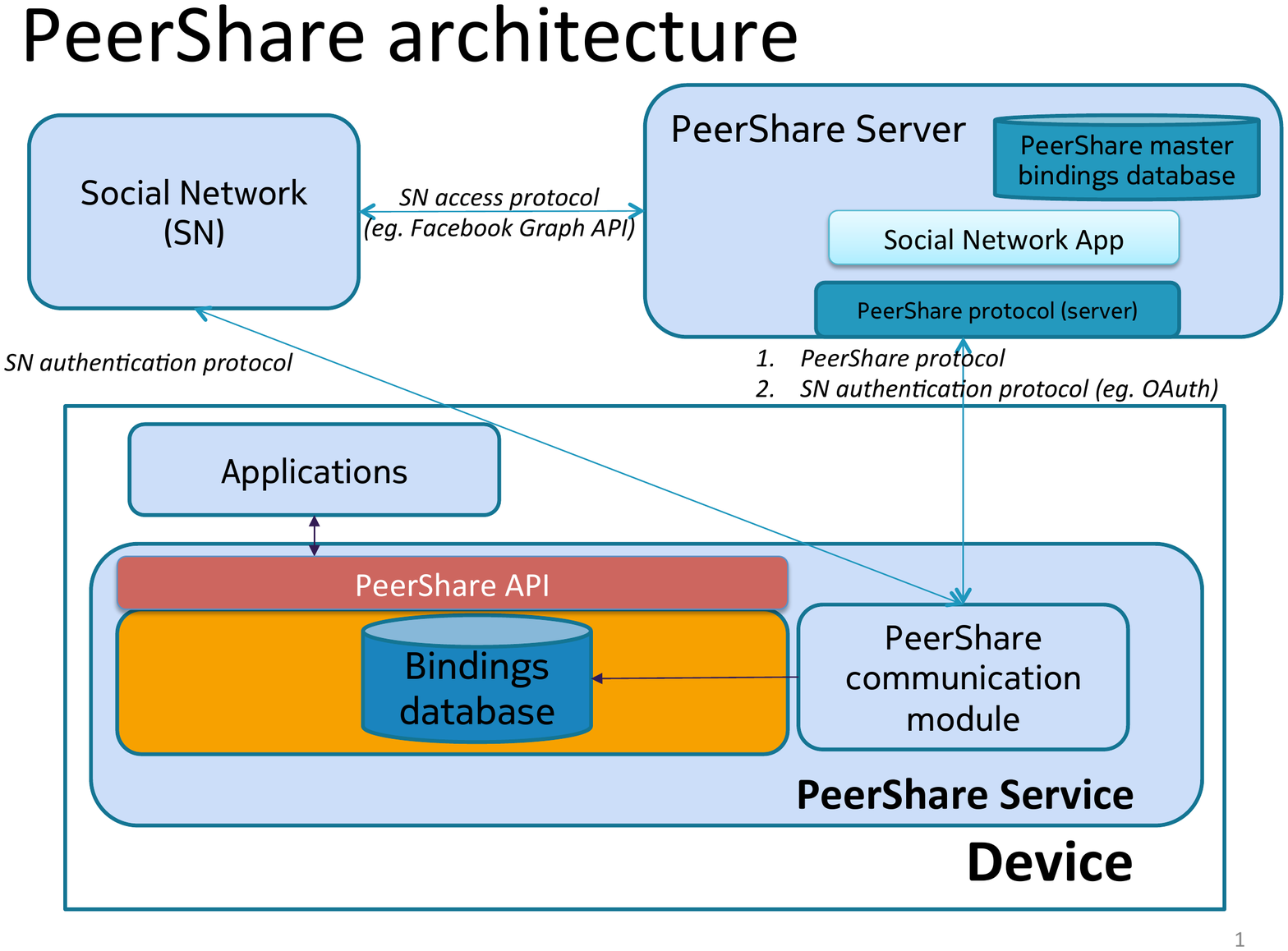}
    \caption{\peershare architecture.}
	\label{system_overview}
\end{figure*}

\subsection{Security requirements}
Communication channel protection is required to prevent a man-in-the-middle attack. The threat of unauthorized usage motivates usage of server authentication, mobile application authentication, user authentication and application access control.

\section{System design}
\label{:design}
The \peershare system allows for creation, storage and distribution of application specific data securely. The system consists of two main components: (1) \peershare Service, and (2) \peershare Server, which are described below. Figure \ref{system_overview} illustrates the system overview.

\subsection{\peershare Service}

\label{:subsec_service}
The \peershare Service is the encapsulation of the \peershare functionality on the mobile device. It is exposed to other applications via the \peershare API described later in the section. It is also responsible for communication with the \peershare Server, which is described in the section \ref{:subsec_protocols}, along with the protocol between the client and the server. 

The heart of the service is internal database that stores data together with their mappings to social identities, which are obtained from the server. The database security is guaranteed by the mobile platform security. Data are bound to social identities by means of social network authentication. 

The service stores a data item inside the AppData data structure which is described by the following attributes:
\begin{enumerate}
\item \textbf{Data type} binds item to particular type (e.g. Bluetooth MAC address), or application (e.g. SCAMPI identifier). Please refer to section \ref{:usecase} for more details. 
\item \textbf{Data value} is the actual value of the item. For instance, it can be a SCAMPI identifier, or Bluetooth MAC address, etc.
\item \textbf{Data description attributes}
\begin{itemize}
\item \textbf{Data algorithm} describes sequence of data procession from its original form to the stored one (e.g. if MAC address is stored as SHA-1 hash, data algorithm value is set to \emph{SHA-1}),
\item \textbf{Specificity} defines if data is device specific (i.e., it is bound to a device like MAC address and multiple values on different devices can be assigned to one user), or it is user specific (i.e., only one data instance can be created for the user. For example, bearer token used to detect presence of your friend),
\item \textbf{Sensitivity} states if data are public or private. For public data (like public key), data authenticity is guaranteed, whereas for private items, data secrecy is additionally guaranteed,
\item \textbf{Binding type} defines two types: \emph{owner asserted binding} and \emph{user asserted one},
\item \textbf{Description} is a human readable text that describes what information the item contains.
\end{itemize}
\item \textbf{Data sharing policy} specifies with whom the item is to be shared. For example, sharing policy may include all my friends in the social network, or just a subset of them.
\item \textbf{Timestamps}
\begin{itemize}
\item \textbf{Data creation timestamp} describes when the data item has been created,
\item \textbf{Data expiry timestamp} specifies when the validity of the data item expires.
\end{itemize}
\item \textbf{Social information}
\begin{itemize}
\item \textbf{Data owner social identifier} is the social ID of the user inside the social network (e.g. user ID in Facebook),
\item \textbf{Data owner social name} is the standard name used by the user in the social network (e.g. John Smith),
\item \textbf{Data owner social network} specifies which social network is used to bind data to particular person (e.g. Facebook).
\end{itemize}
\item \textbf{Creator application identifier}
\begin{itemize}
\item \textbf{Mobile platform identifier} (e.g. Android, iOS, Windows Phone),
\item \textbf{Platform-specific application identifier} specifies minimal set of elements allowing for unique identification (within a given mobile platform) of an application that has created the data (e.g. package name and developer key for Android).
\end{itemize}
\end{enumerate}
Table \ref{data_summary} presents the summary of the most important AppData attributes.

Furthermore, the service provides application level data access control which guarantees that only authorized applications can modify or delete existing data. Any application can create data that it intends to share using the system. During the initial data upload process, the service records the calling application platform specific identifier (e.g. pair of Android package name and developer key) and appends it to the created data. As a result, if an application wants to modify or delete existing data, the service learns the calling application identifier and verifies it against the application identifier recorded in the initial upload process. 

The final responsibility of the service is a periodic update of the database state. The service has implemented a timer mechanism that fires every six hours. Upon it, the service tries to connect to the server and fetch all eligible data from it. If the service is not running in the instant of timer expiry, the refresh operation is postponed until the service is started again. The fact that the service is not running also means that there is no third party application that needs access to data, thus postponing of database update does not introduce any limitations to the service. 

\begin{table*}[ht!]
\caption{Summary of \peershare data attributes}
\label{data_summary}
\centering
\begin{tabular}{| c | m{0.65\textwidth} |}
\hline
\textbf{Data attribute} & \textbf{Description} \\ \hline
Data type & Mapping of data to particular type, or application \\ \hline
Data value & Actual value of data \\ \hline
Data description attributes & Provides more detailed description of a data item by indicating algorithm used for its creation, specificity, sensitivity and binding type \\ \hline
Data sharing policy & Specifies sharing policy for data \\ \hline
Timestamps & Indicates timestamps for data creation and its validity\\ \hline
Social information & Social information that is bound to data \\ \hline
Creator application identifier & Identifies mobile platform and application that owns a data item \\ \hline
\end{tabular}
\end{table*}

The \peershare API is the second part of the mobile application. It provides the interface for third-party applications for creation, modification and removal of data that an application wishes to share with other application users. The most important methods are described below:
\begin{description}
\item[long addData(AppData data)] \hfill \\
When an application wishes to share new data through the \peershare system, it should use this method. It stores application data and uploads it to the \peershare Server as soon as the network connectivity to the server is established. The method returns object identifier which the application should use in the future to modify, or delete this data item.
\item[int updateData(long objectID, AppData data)] \hfill \\
This method is intended for modification of already existing data. The object ID obtained in the \emph{addNewSecret} method identifies the data item to update. 
\item[long removeData(long objectID)] \hfill \\
This method deletes existing data. It requires the object ID to identify which data is to be removed. The object ID obtained in the \emph{addNewData} method identifies the data item to update. 
\item[Other methods] \hfill \\
The application interface defines also other methods that can be used for retrieval of specific data (e.g. \emph{getSharedDataDetail}), finding out social information related to the currently logged in social network user (i.e., \emph{getMySocialData}), and learning available data access control policies (i.e., \emph{getACLPolicies}).

\end{description}

\subsection{\peershare Server}

The \peershare Server is the trusted entity that is primarily responsible for secure storage of data. Every data item is bound to a social identifier of the user (e.g. Facebook user ID) who has created the item. The server authenticates users by requiring them to provide a valid social network user access token and verifying its correctness through interaction with the social network server. 

The second crucial responsibility of the server is enforcement of access control policies for stored data. The system allows users to specify who is eligible to access stored data by allowing them to state the sharing policy from all available social network user lists of the user. The server queries the social network server for custom friend lists created in the social network by a user. Such lists are returned to the \peershare service and can be further accessed by other applications to allow them specify the sharing policy. On creating a new data item, or updating sharing policy for the existing item, the server queries the social network server to obtain the list of social user IDs applicable to download particular data item. If an application uploading a new data item does not specify its sharing policy, the data item is by default shared among all user's friends.

Furthermore, the server updates lists of users assigned to particular friend list in case of their modification by means of the realtime updates provided by the social network. If the social network does not have this functionality, the server must regularly poll the social network server to learn about changes in the friend lists.

\begin{figure}[hb!]
	\centering
	\subfigure[Screen presenting data owned by a user]{ \label{my_data} \fbox{\includegraphics
	[scale=0.185] {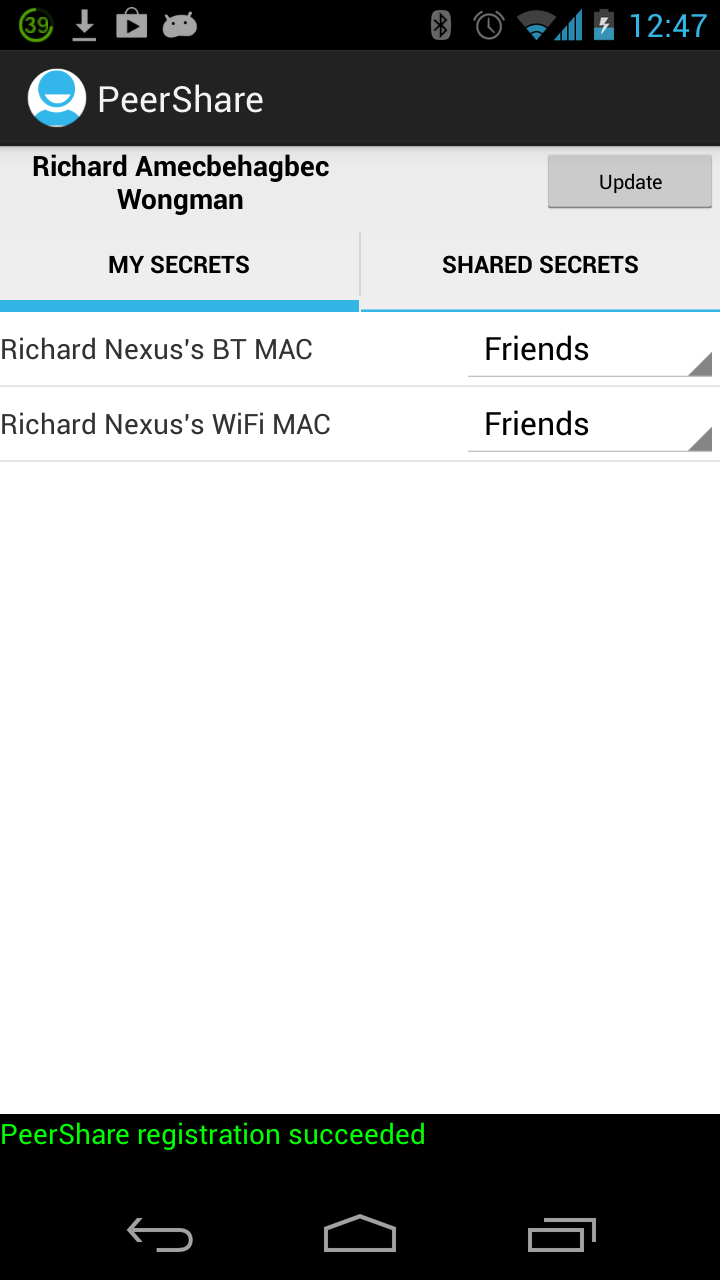}} }
	\subfigure[Screen presenting data shared with a user]{ \fbox{\includegraphics
	[scale=0.185] {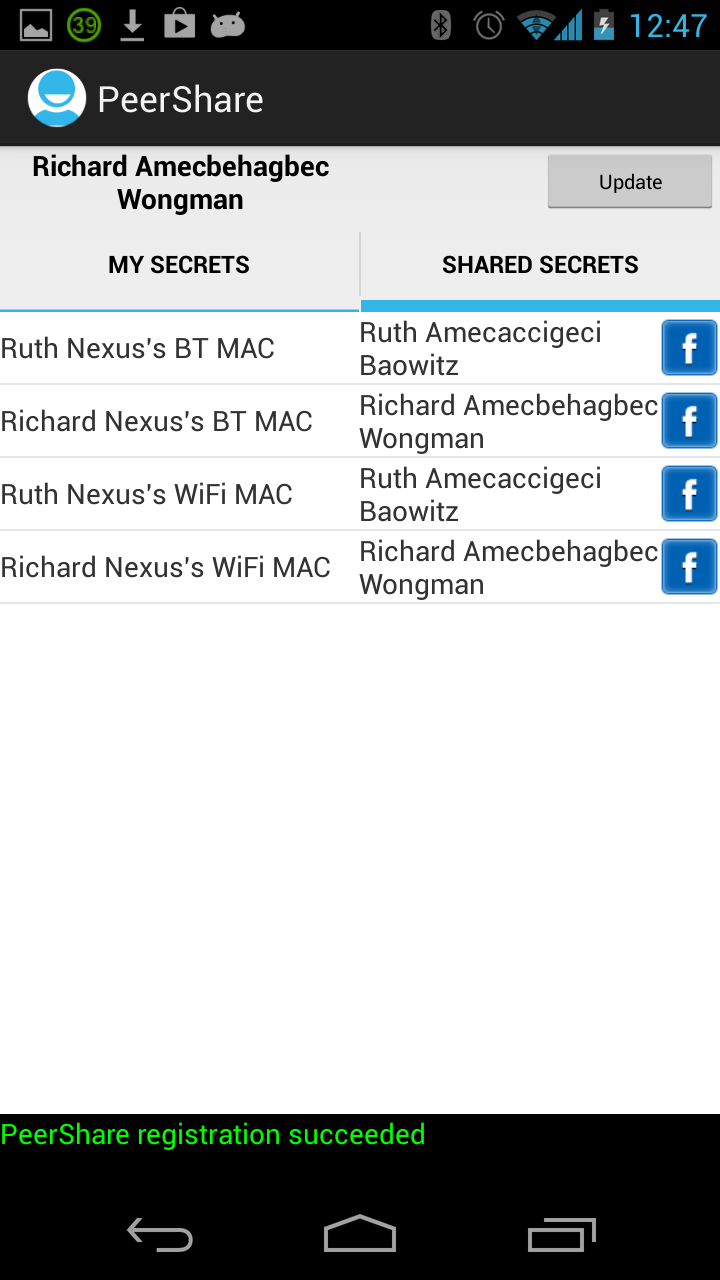}} }
    	\caption{Screenshot of the \peershare graphical user interface for Android}
	\label{implementation_screenshot}
\end{figure}

Finally, users are allowed to overwrite data sharing policy specified by an application via the web interface (see figure~\ref{my_data}). In such situation, sharing policy specified by the user is updated on the server that is going to distribute updated item according to the new sharing policy.

\subsection{\peershare Protocol}
\label{:subsec_protocols}

The \peershare service communicates with the server through the JSON encoded protocol that runs on top of standard HTTPS protocol. It involves following operations: user registration and unregistration, data upload and update, and data download. Because data exchanged between the service and the server are sensitive, communication security is guaranteed by the TLS layer. Furthermore, to protect against fake social network application attack, each request includes user access token of the social network, whose validity is verified by the server. 

In the \textbf{\emph{REGISTER}} method, the user registers for the service by informing the server about his/her social information. In response, the server generates (or finds if the user is not a new one) user's \peershare identifier that is needed in all subsequent transactions with the server to uniquely identify the correct person. The \peershare ID is necessary to properly correlate possible multiple social identities of the same person. This may happen if someone uses more than one social network in the \peershare system (e.g. Facebook and Twitter).

In the \textbf{\emph{UPLOAD}} method, the service sends to the server all data items which have been added to the database on the user's device, but have not been uploaded on the \peershare server. The message contains also the \peershare ID to map uploaded data to a specific user. Content of each data item is consistent with data description provided in the section \ref{:subsec_service}. In response to the \emph{UPLOAD} request, the server sends an array of object IDs that are later used to modify or delete every data item from the server database.

The \textbf{\emph{UPDATE}} method is very similar to \emph{UPLOAD}. The only difference is that it is not adding any new data on the server, but only updating existing ones. The object ID returned in the \emph{UPLOAD} operation is needed to properly identify the item on the server to modify. If the service wants to update a non-existing item (i.e., the one that has already been deleted by the user), the server ignores the request to do it, and sends back in response notification that the data item does not exist and should be removed from the local database.

The \textbf{\emph{DOWNLOAD}} method allows the service to fetch all data items that the registered user is eligible to obtain. It requires the service to provide user's \peershare ID to correlate the request with the correct user. In response, the server returns an array of data items that contain detailed information about each item in a format similar to the one used in the \emph{UPLOAD} or \emph{UPDATE} request. The only difference is that personal information (i.e., sharing policy, and object ID) is not included unless the downloading user is the owner of the item.

The \textbf{\emph{DELETE}} method is used to erase old and no longer needed data from the server. Similarly to all other methods, it must include the \peershare ID to correctly correlate user with data. In addition, it also contains an array of object identifiers that are to be deleted. In response, the server sends just status information that is either \emph{OK} if there are no errors, or is an error message.

\textbf{\emph{UNREGISTER}} is the final method defined in the protocol. It is used to unregister the user from the \peershare and delete all data associated with the user. In the request, the \peershare ID together with the social network identifiers are provided. The server responds with \emph{OK} status, or an error message if it fails to unregister the user.

Table \ref{protocol_summary} provides a short summary of all protocol methods.

\begin{table*}[ht!]
\caption{Summary of \peershare protocol methods}
\label{protocol_summary}
\centering
\begin{tabular}{ | c | m{0.7\textwidth} | }
\hline
\textbf{Protocol method} & \textbf{Description} \\ \hline
REGISTER & Registers user to the system. Creates unique \peershare identifier for the user to utilise later in every interaction with the server.\\ \hline
UPLOAD & Uploads new data to the server. \\ \hline
UPDATE & Modifies already existing data with updated values. \\ \hline
DOWNLOAD & Fetches all data shared with the user from the server to the mobile device. \\ \hline
DELETE & Deletes no longer needed data from the server. \\ \hline
UNREGISTER & Erases all user data from the server when he/she does not want to use the system anymore. \\ \hline
\end{tabular}
\end{table*}

\subsection{Implementation}
Our server implementation consists of a couple of PHP scripts, PostgreSQL database and the Facebook PHP SDK. Currently the server supports only Facebook as the social network to authenticate with, but the architecture is generic enough, so that in the future it can be easily extended to support other social networks. Finally, the server takes advantage of Facebook Realtime Updates functionality to learn about modifications of user's lists.

The service implementation is more complex, as it includes the communication module as well as API for third party applications. Currently we have the Android implementation of the client package. The service is implemented as a standard Android background service that runs as an independent process. It contains internal SQLite database, where \peershare data are stored. Applications using the service bind to it through the AIDL interface. SSO user authentication is currently provided by native Facebook library. If multiple social networks are available in the future, native libraries providing SSO authentication will also be added. The service uses also Binder interface functionality to learn about service calling application identifiers. It allows matching calling applications with data they create, which is critical to provide application access control.

\subsection{Deployment and performance considerations}

The \peershare system has up to now been successfully deployed in three projects, as described in the section~\ref{:usecase}.

To evaluate performance of the \peershare, we have tested average time needed for upload and download of data in the WiFi network that uses ADSL connection. In our test scenarios, a user uploads 1 data item, and downloads 5 data items, as 5 friends share sample data in a test application. Average upload time measured in 30 runs is 2.02 seconds with standard deviation of 1.33 seconds. Download operation performance is more stable, as average time is 1.18 seconds with standard deviation of 0.12 seconds. To compare these numbers with standard web browsing activities, we conducted similar experiments for downloading mobile Facebook web page. Average download time for Facebook webpage measured in 30 runs is 1.50 seconds with 0.21 seconds of standard deviation.

Furthermore, \peershare has been designed to allow multiple applications use the same server. However, if application performance is limited due to the server scalability, each application developer may decide to run its own server. Such a solution is further discussed in the section~\ref{:minimizing_trust} describing possibility of minimizing the need of trust for the \peershare Server. 

\section{Security considerations}
\label{:validation}
In this section, we present our security analysis showing that the security requirements presented in the section \ref{:requirements} are fulfilled.

\subsection{Channel protection}

In order to guarantee channel protection, \peershare Protocol is
executed over a secure (i.e., confidential and mutually authenticated)
channel.  The user is authenticated by the OAuth protocol via the
native Android Facebook application. Therefore the system relies on
correct behaviour of the native Android Facebook application.

\peershare Server is authenticated via a TLS certificate, 
We use a form of ``certificate pinning'' by embedding the TLS server
certificate of \peershare Server in the client implementation.  This
protects against a rogue server from masquerading as \peershare Server
even if the rogue server has succesfully obtained a certificate for
its TLS keypair from one of the tens of Certification Authorities that
are normally trusted for TLS.  If there are many \peershare Servers,
then instead of hardwiring the TLS server certificate, we can use
standard certificate pinning~\cite{draft.pinning}.

\subsection{User and application authentication}
User authentication is obtained through the native Facebook Android library. Prior to invoking any interaction with the server, the service asks the native Facebook application (through the library interface) for a valid access token associated with the authenticated user. Such a valid token must be included in every message exchanged with the server. The \peershare Server uses the Facebook graph API token debug tool to examine its validity by checking the following:
\begin{itemize}
\item does application identifier encoded inside the token correspond to the \peershare Facebook application identifier
\item does user identifier encoded inside the token correspond to the social identifier included in the sent message
\end{itemize}
The former checking protects the server against allowing a fake Facebook application to modify data on the server. The latter one prevents other users from modifying or deleting data that do not belong to them. Only if both above conditions are fulfilled, the \peershare server proceeds with the request. Otherwise, it responds to the sender with authentication error.

\textbf{User access control}
The user access control must guarantee that only eligible users are able to obtain data from the \peershare server and that only the user that has created a particular data item can modify or delete it. Correct data distribution is secured by the server that learns data sharing policy from the request to store/update the data item. For each created/updated sharing policy, the server interacts with the Facebook graph API to fetch the list of social user IDs associated with the given policy. Having obtained such list, the server stores information about social IDs eligible to download particular data item. 

The second problem is resolved on the device side. Whenever a third party application makes a request to modify or delete the particular data item, the \peershare service checks in the local database if the item has been created by the user trying to modify it. If this is true, the service grants application permission to edit or remove the item. Otherwise, it denies application access to the given item.

\textbf{Application access control}
As multiple applications on one mobile device can use the \peershare system, there is a threat that a malicious application using the system can modify or delete a data item that does not belong to it. In order to protect against this threat, the \peershare service has built-in application level access control enforcement. When an application creates a new data item, and wants to have it distributed through the \peershare system, the service tries to infer the platform specific calling application identifier and appends it to the uploaded data. For Android operating system, the application identifier is a tuple of package name and developer public key that can be obtained through the Binder interface. On subsequent requests to update/delete the data, the service again obtains the identifier of the calling application and compares it with the one associated with object as the creator. In the Android operating system, this function is performed by verification if package signatures match. Unfortunately, some operating system may not permit for the service to infer the calling application identifier. In such case, the caller must explicitly specify the application identifier.

\subsection{Minimizing the need to trust \peershare Server}
\label{:minimizing_trust}

Since \peershare Server has access to all the sensitive data, it
needs to be trusted by all participants. 
This is a rather strong assumption.  There are two ways to reduce the
extent to which \peershare Server needs to be trusted:

\noindent\textbf{Use of trusted hardware}:
If \peershare Server is equipped with a hardware security module (HSM) like
the Trusted Platform Module
(TPM)\footnote{\url{http://www.trustedcomputinggroup.org/resources/tpm_main_specification}},
then \peershare server database can be encrypted using a
HSM-resident key.  HSM will decrypt the plaintext and make it
available to a process if and only if the host computer is in the
correct configuration (i.e., running the correct \peershare
Sever software). An attacker will have to subvert \peershare Server process
at runtime.  If the client devices also have a hardware-based trusted
execution environment (like On-board
Credentials~\cite{DBLP:conf/ccs/KostiainenEAR09}, then the server
HSM can encrypt the sensitive data so that it is accesible only within
the client TEE, thereby not exposing it to the \peershare Server at
all.

\noindent\textbf{Application-specific \peershare Server}:
Although we designed \peershare in such a way that multiple
application developers could use the same \peershare Server, in
practice, each application developer could decide to host her own
independent \peershare Server.  This would still allow the benefit of
developer ease of use because developers can re-use our \peershare
implementation, without asking all developers to trust the same
server.

\section{Related work}
\label{:related}
The concept of data sharing with social networks support is present also in other works. SocialKeys~\cite{socialkeys} project proposes the idea of distributing public keys via social networks. \peershare extends this concept to: various types of data and multiple applications. 

Backes et al.~\cite{DBLP:conf/ndss/BackesMP11} presents a generic cryptographic framework that allows social relations establishment and resource sharing with user anonymity, secrecy of resources, privacy of social relations and access control secured. Unlike \peershare that uses social network specified sharing policies, it requires users to explicitly establish social relationships with other users which makes it less intuitive for users in real deployments.

Baden et al. has implemented Persona~\cite{DBLP:conf/sigcomm/BadenBSBS09}, a distributed social network with distributed data storage. It provides data access control by employing a combination of traditional public key cryptography and attribute-based encryption (ABE) scheme that involves more complex key management. Safebook~\cite{Cutillo:2009:SPO:2288675.2294014} is the implementation of the distributed social network that improves privacy protection mechanisms in comparison to other existing social networks. Improved privacy results from taking advantage of cooperation between users inside a peer-to-peer overlay network, named matryoshka, and trust relations among users to achieve integrity and privacy properties. Unlike concepts presented in these works, our goal is not to build a new social networks, but to make use of existing social networks, as one of the requirements of the system is its deployability. Obviously, if any of these social networks proves to be successful, we are interested in taking advantage of their security and privacy mechanisms in \peershare.

Jahid et al. has implemented DECENT~\cite{DBLP:conf/percom/JahidNMBK12} that is a decentralized social network system providing confidentiality and integrity of data that due to cryptographic mechanisms can be stored in untrusted nodes. Unlike our system, it requires user to explicitly specify data sharing policy and build their social relationships, thus it is more difficult to use in real life than \peershare.

\section{Conclusion}
\label{:conclusion}
We have described the design and implementation of \peershare, the system that allows users for secure and privacy preserving data distribution system. It implements a generic framework for data distribution that can be used by different applications and provides different levels of security guarantees. Furthermore, the system enhances general usability for end-users and application developers by using existing and popular social network mechanisms for the user authentication and data distribution inside a specific social context.

\peershare has been used by three different applications. We plan to make it available to other application developers.

\bibliography{ref}
\bibliographystyle{plain}

\end{document}